\input{aipcheck}

\documentclass[
    ,final            
  ]
  {aipproc}

\layoutstyle{6x9}

\usepackage{amsmath,amssymb}

\begin{document}

\title{\vspace*{-2cm} \hspace{10cm}{\small \hfill{CPT/10/46}}\\[-0.2cm]\hspace{10cm}{\small \hfill{IPPP/10/23}}\\[1cm]
Light particles\\ --\\ A window to
fundamental physics}


\author{Joerg Jaeckel}{
  address={Institute for Particle Physics Phenomenology, Durham University, Durham DH1 3LE, UK}
}



\begin{abstract}
In these proceedings we illustrate that light, very weakly
interacting particles can arise naturally from physics which is fundamentally connected to very high energy scales.
Searching for them therefore may give us interesting new insights into the structure of fundamental physics.
Prime examples are the axion, and more general axion-like particles, as well as hidden sector photons and matter charged under them.
\end{abstract}

\maketitle


\section{Introduction - Hope for light particles}
Over the years both theoretical as well as experimental evidence has accumulated that strongly suggests the existence of physics beyond the current standard
model of particle physics (SM).
A particular focus of attention has been the TeV scale. Indeed there is a lot of circumstantial evidence that exploring the TeV scale
will bring decisive insights into fundamental
questions such as the origin of particle masses, the nature of dark matter in the universe, and
the unification of all forces, including gravity.
Indeed, most proposals to embed the Standard Model of particle physics into a more general,
unified framework, notably the ones based on string theory or its low energy incarnations,
supergravity and supersymmetry, predict new heavy, $m\gtrsim 100$~GeV,
particles which may be searched for at TeV colliders.
The Large Hadron Collider currently starting up at CERN will test many of these ideas for such physics beyond
the standard model (BSM) and hopefully
will provide us with a wealth of new information.

In this note\footnote{This note is a very rough and heuristic
overview. More details and references can be found in
\cite{Jaeckel:2010ni}.} we argue (mostly in a heuristic and
qualitative way) that there is also a very good motivation to search
for new physics in low energy experiments that can provide us with
powerful complementary information on currently open questions and
in particular on how the standard model is embedded into a more
fundamental theory.

The first simple but also very compelling reason for new physics at low energy scales is purely phenomenological:
we have already discovered a variety of effects that seem to be connected to low energy scales:
\begin{itemize}
\item{Neutrinos have masses of the order of meV.}
\item{The energy density of dark energy is of the order of $(\rm meV)^4$.}
\item{The total energy density of the universe (of which dark energy is about 70 \%) is of the order of $(\rm meV)^{4}$.}
\end{itemize}
Although this could be coincidental it definitely suggests that exploring the sub-eV regime of particle physics is a worthwhile enterprize.

Phenomenologically new, light particles must be extremely weakly coupled to the SM in order to have avoided detection until today.
This brings us to a second perhaps even more compelling reason to explore the low energy frontier. {\emph{Small masses and weak couplings}} might be inherently
connected to physics occurring at very {\emph{high energy scales}}.
Let us demonstrate this by using the axion as an example.

The QCD axion can arise as the (pseudo-)Goldstone boson of a
spontaneously broken U(1) Peccei-Quinn
symmetry\footnote{This symmetry is introduced to explain why the strong interactions do not violate CP and consequently
why no electric dipole moment of the neutron is observed. In this sense the axion is also a prime example of a new light
particle that is directly motivated by a ``bottom-up'' solution of an existing phenomenological problem in the SM.}~\cite{Peccei:1977hh}. Consequently it couples only via
higher dimensional operators (or derivative couplings). A typical
coupling is the coupling of the axion to two
photons, 
\begin{equation}
{\mathcal L}_{a\gamma\gamma}=-\frac{1}{4}g_{a}\,a\,F_{\mu\nu}\tilde{F}^{\mu\nu}.
\end{equation}
The coupling constant $g_{a}$ has dimensions $1/mass$ and is given by
\begin{eqnarray}
        { g_{a}} = \frac{\alpha}{2\pi f_a}
\left( {\frac{2}{3}\,\frac{m_u+4 m_d}{m_u+m_d} - s    }\right)
\sim 10^{-13}\ {\rm GeV}^{-1}          \left(
         \frac{10^{10}\, {\rm GeV}}{f_a}\right),
         \label{axionphotoncoupling}
\end{eqnarray}
where $f_{a}$ is the scale at which the Peccei-Quinn symmetry is spontaneously broken, $m_{u,d}$ are the masses of the up and down quarks and $s$ is a model
dependent constant of order one.
Now, the scale of new physics is given by $f_{a}$ (everything else is electroweak physics).
The crucial feature is that the coupling is suppressed by a large axion decay constant. Therefore, probing
small couplings means we are actually probing very large energy scales.

For low energy probes it is also important that the mass of the interesting particles is small. This is also
true for the axion. Its mass is given by
\begin{eqnarray}
m_a =
         \frac{m_\pi f_\pi}{f_a}\frac{\sqrt{m_u m_d}}{m_u+m_d}\simeq { 0.6\,  {\rm meV}}
         \times
         \left(
         \frac{10^{10}\, {\rm GeV}}{f_a}\right),
         \label{axionmass}
\end{eqnarray}
where $m_{\pi}, f_{\pi}$ are the pion mass and decay constant.
Again the mass is suppressed by the large energy scale $f_{a}$, as befitting a pseudo-Goldstone boson.
Probing axions in the meV regime actually means probing Peccei-Quinn scales of the order of $10^{10}$~GeV -- a truly
enormous scale, a factor $10^7$ above the scale probed by LHC.

This (well-known) example shows that indeed small coupling and very
low masses can be connected to new physics occurring at very high
energy scales. In the next section we will explore how this (and
other mechanisms generating small masses and couplings) could be
realized in extensions of the standard model based on string theory.

\section{WISPs from string theory}
In this section we will give a rough picture of how (very) weakly interacting sub-eV particles (WISPs) could
arise in string theory.
We will be mainly concerned with two types of WISPs:
\begin{itemize}
\item{
Axion-like particles, i.e. scalar
or pseudoscalar particles coupled to two photons. These particles
are very similar to the QCD axion, but we do not insist on the
relation between the coupling to two photons and the mass of the
particle predicted in models of the QCD axion\footnote{And, of
course, they do not necessarily have to solve the strong CP
problem.}  (cf. Eqs.~\eqref{axionphotoncoupling}, \eqref{axionmass}).}
\item{Hidden sector U(1) gauge bosons and hidden sector matter charged under them.}
\end{itemize}

String theory is a top-down approach for physics beyond the standard model. It ambitiously tries to unify the SM with gravity and in order to do so
replaces point particles (a basic ingredient of quantum field theory) with one-dimensional extended objects, so-called strings.
(For concreteness we will focus in the following mainly on the mechanisms present in so-called type II string theories.)

\begin{figure}[t!]
\centerline{\includegraphics[width=1.0\textwidth]{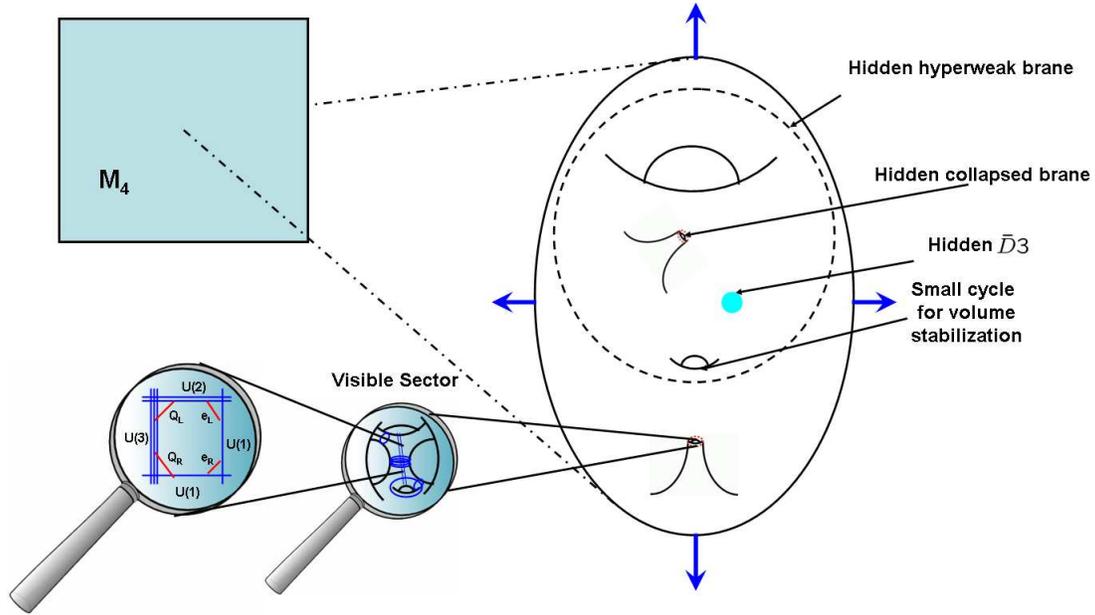}}
\caption{In compactifications of type II string theories the
Standard Model is locally realized by a stack of space-time filling D-branes wrapping
topologically non-trivial submanifolds in the compact dimensions. In
general, there can and often must also be hidden sectors localized at different
places. They can arise from branes of different dimension (D3 or D7
branes) which can be either of large extent or localized at
singularities. Light visible and hidden matter particles arise from
strings located at intersection loci and stretching between brane
stacks. Gauge bosons (not shown) correspond to strings starting and ending on the same stack of branes.
The blue arrows denote potential shape and size deformations of the compactified extra dimensions corresponding to
scalar fields in 4 dimensions.
}\label{Fig:type_ii_comp}
\end{figure}

At the moment, however, it is not yet possible to derive low energy physics (in this case this includes physics at the TeV scale) directly from string theory.
Nevertheless, string theory gives us information on what type of structures and features are expected including a wide variety of constraints
on what is allowed and what not. Using these we can now construct low energy models and study their properties. An example of such a model is
shown in Fig.~\ref{Fig:type_ii_comp}.

%

\subsection{Axion-like particles}
We can now look at such a model and see what features may give rise to phenomena observable at very low energies. The first important
feature is the existence of extra space dimensions. String theory needs extra dimension for consistency. More precisely, string theory
typically lives in 9+1 dimensions.
Naively this is in obvious conflict with observation and we have to make them
invisible by wrapping them up on a very small length scale, i.e. we have to compactify them.
However, compactification leaves traces in the form of new particles.
After compactification the size and shape of the extra dimension can still change. Naively one can imagine that they start to vibrate (cf. blue
arrows in Fig.~\ref{Fig:type_ii_comp}) around a preferred form and size.
These size and shape deformations correspond to four dimensional fields/particles, so-called moduli.

%
%

These particles are typically light (often even too light, i.e. they are ruled out by fifth force experiments~\cite{Adelberger:2009zz}).
Moreover, their masses and couplings are directly related to the scale of compactification and the fundamental scale, in this case
the string scale $M_{s}$. Therefore, testing these particles could give us crucial insight into the structure of the compactification and the
underlying fundamental scales.

Beyond their very existence we can even argue that these particles can easily be coupled to the SM and even electromagnetism. Moreover, each
scalar particle naturally comes with a pseudoscalar counterpart that couples like an axion\footnote{Here, we closely follow the argument given in~\cite{Conlon}.}.
The Lagrangian of a gauge theory typically has the form,
\begin{equation}
{\mathcal L}= -\frac{1}{4g^2} F^2 - \frac{\theta}{32\pi^2}F\tilde{F}.
\end{equation}
In an ordinary field theory $g$ and $\theta$ are independent parameters.
Now, we can make crucial use of two features of string theory. First of all, in string theory there is only one parameter, all other
parameters should be given in terms of this parameter and vacuum expectation values of fields, in particular the fields describing
the size and shape of the extra dimensions. Therefore we have $g(\Phi)$ and $\theta(\Phi)$. The second crucial feature
is supersymmetry. Supersymmetry is a (nearly) necessary property to make string theory consistent.
Supersymmetry forces the functions for $g$ and $\theta$ to be related,
\begin{equation}
{\mathcal L}={\rm Re}[f(\Phi)]F^2+{\rm Im}[f(\Phi)] F\tilde{F}.
\end{equation}
Indeed we can read off that the first term looks like a scalar particle coupled to two gauge bosons (e.g. photons) whereas the the second
part is a pseudoscalar coupling to two gauge bosons, the trademark coupling of an axion-like particle. So overall we have couplings suitable for scalar as well
as pseudoscalar axion like particles.

Masses and couplings of these axion-like particles are typically directly connected to the fundamental mass scale $M_{s}$ and the
volume of the compactification~\cite{Conlon:2006tq},
\begin{equation}
\frac{1}{g_{\rm ALP}}\sim f_{\rm ALP}\sim \frac{M_{P}}{{\rm
Vol}^x}\sim M_{s}\left(\frac{M_{s}}{M_{P}}\right)^y,\quad\quad
m_{\rm  ALP}\sim \frac{\Lambda^2}{f_{\rm ALP}}, 
\end{equation}
where $\Lambda$ is a model-dependent scale of explicit symmetry breaking, and $x$ and $y$ are model-dependent numbers.
Overall a wide range of masses and couplings is possible, but via their connection to the fundamental scale these particles may nevertheless give us
interesting information on the structure of the underlying theory.

\subsection{Hidden sector gauge bosons and matter}
In addition to the existence of extra space dimensions, another generic feature of string theory may give rise to new, possibly light particles
that interact only very weakly with the SM: Hidden sectors.

As we can see from Fig.~\ref{Fig:type_ii_comp} particles correspond to strings attached to space-time filling membranes.
Very roughly speaking a stack of N membranes sitting on top of each other gives rise to a U(N) gauge group, and the gauge bosons are the strings stretching from this stack
back to itself. Matter particles are strings stretching between different stacks of branes -- a string stretching from an N-stack to an M-stack transforms
as a bifundamental under the U(N) and U(M) gauge groups.
We can construct the SM (cf., e.g.,~\cite{Aldazabal:2000sa}) by inserting an appropriate configuration of stacks at some place in our compactified space (a suitable place is typically some singularity
or cycle) as shown in Fig.~\ref{Fig:type_ii_comp}.

However, constructing the SM in this way is typically not consistent, the resulting setup has anomalies and other potentially undesirable features.
In order to cure these one typically has to introduce other stacks of branes to cancel anomalies etc.. These generate a whole new sector of gauge fields
and corresponding matter particles. In order to make them weakly interacting with the SM particles we can (try) to hide them by placing them at a large distance in the extra
dimensions away from the SM branes. Roughly the idea is as follows, if the distance between to stacks of branes is large, strings stretched between the two stacks
are very heavy and the interactions mediated by these particles very suppressed.
The extra branes become ``hidden sectors''.

In scenarios, where the volume of the extra dimensions is very large, we have an additional way to hide extra sectors (at least to some degree). We
can drastically increase the extension of their branes in some of the extra dimensions (cf. the dashed line in Fig.~\ref{Fig:type_ii_comp}).
Naively this dilutes the gauge coupling making it ``hyperweak''~\cite{Burgess:2008ri}.

As we have seen stacks of branes typically correspond to U(N) gauge groups. In particular they contain U(1) factors. Now, the special feature
of U(1)s is that, in field theory, they can interact with other U(1)s directly via a renormalizable dimension four term in the Lagrangian, a so-called
kinetic mixing~\cite{Holdom:1985ag}\footnote{Actually a similar mixing is possible via a mixing $\theta$-term, called magnetic mixing~\cite{Bruemmer:2009ky}.},
\begin{equation}
\label{kinetic}
{\mathcal L}=-\frac{1}{4}F_{\mu\nu}F^{\mu\nu}-\frac{1}{4}X_{\mu\nu}X^{\mu\nu}+\frac{\chi}{2}F_{\mu\nu}X^{\mu\nu}.
\end{equation}
Here, $F$ and $X$ are the field strength tensors of the two U(1)s. In order to connect to potentially observable effects let us take
one of the U(1)s, say $F$, to be our ordinary electromagnetic U(1) and the other, $X$, to be one of the hidden sector U(1)s.

The fact that an interaction is possible via a dimension four operator has dramatic consequences. Let us imagine that the kinetic mixing
is generated by the effects of some very heavy particles interacting with both branes. However, in contrast to higher dimensional operators,
the {\emph{dimensionless}} mixing parameter
$\chi$ generated is not naturally suppressed by powers of the mass of these particles.
Therefore kinetic mixing provides a natural window to probe these hidden sectors. (It should be noted however, that the naive dimensional
arguments can fail and we may get some suppression by the distance, i.e. volume~\cite{Lust:2003ky}. Moreover, additional symmetries may also lead to a smaller
or even vanishing kinetic mixing~\cite{Dienes:1996zr,Lust:2003ky}.)

Let us briefly discuss, how the kinetic mixing can become observable. If we have matter charged under the hidden U(1), one finds by diagonalizing the
kinetic term, Eq.~\eqref{kinetic}, that the hidden matter particles obtain an electric charge $\sim \chi$ under our ordinary electromagnetic U(1)~\cite{Holdom:1985ag}.
But, even without additional hidden matter fields the kinetic mixing is observable if the extra U(1) has a (small) mass, we then
observe photon -- hidden photon oscillations, analog to neutrino oscillations~\cite{Okun:1982xi}.
Both effects may allow us to detect very small kinetic mixings, in some mass regions currently down
to $10^{-14}$ (see~\cite{Davidson:2000hf,Jaeckel:2010ni} and Refs. therein).

In total, we again find a, somewhat model-dependent, prediction for the masses and kinetic mixing which is, however, directly connected
to the fundamental parameters and the details of the compactification~\cite{Lust:2003ky},
\begin{equation}
\chi\sim \frac{g_{\rm vis}g_{\rm hid}}{16\pi^2 ({\rm Vol})^{u}}\sim \frac{2\pi g_{s}}{{\rm Vol}^{v}}\sim \left(\frac{M^{2}_{s}}{M^{2}_{P}}\right)^{w}
\ll 1,\quad
m_{X}^2\sim \frac{g_{s}M^{2}_{s}}{2}\left(\frac{4\pi}{g^2_{s}}\frac{M^{2}_{s}}{M^{2}_{P}}\right)^x
=g_{s}\frac{M^{2}_{s}}{({\rm Vol})^{y}},
\end{equation}
where, $g_{s}$ is the string coupling and $u,v,w,x,y$ are model-dependent numbers. Note, that not only can the interaction be weak,
but also the masses can be volume suppressed, making them potentially
very small.

\section{Conclusions}
Over the last two decades the observation of neutrino masses as well as dark energy has given us intriguing hints
towards the existence of new physics at sub-eV scales.
In this note we have outlined a variety of reasons why the existence of new light particles with very weak interactions
with ordinary matter is very plausible in extensions of the Standard Model and could provide us with unique opportunities to
probe fundamental physics.

(Pseudo-)Goldstone bosons of symmetries spontaneously broken at very high scales, as for example the axion, automatically
have very weak couplings but also very small masses making them accessible to low energy experiments.
In models of fundamental physics, such as string theory, the real and imaginary parts of the fields corresponding to
shape and size changes of the compactified extra dimension give naturally rise to axion-like particles coupled to two gauge bosons. Their interactions
are typically connected to the fundamental scale, such as the string scale.
In addition, we typically have
a wide variety of potentially light hidden sector particles, such as extra U(1) gauge bosons and hidden matter particles which can interact via kinetic or magnetic
mixing with Standard Model particles thereby becoming accessible to low energy experiments.

Both the axion-like particles arising from the deformations of the extra dimensions and the hidden sector particles are inherently connected to the
global features of the extra dimensions. For the former this is obvious, while for the latter one can naively picture them to be ``far away'' in the extra dimension
or, in the case of hyperweakly interacting hidden sectors, stretching through a large volume of the extra dimension.
In this picture, the SM is typically in a relatively localized place in the extra dimension. High energy experiments such as the LHC searching for
new heavy but also relatively strongly interacting particles precisely probe this local structure and its neighborhood.
On the other hand, low energy, high precision experiments searching for very weakly interacting particles may give us complementary information on the global structure.
The price to pay is that we have to rely on the (as we have argued justified) hope that some of the new particles connected to this global structure are light.

Overall, new physics at very high energy scales may not only introduce new heavy particles, but it can also introduce new, very light but also
very weakly interacting particles. The interplay between small couplings, small masses and the high energy scales responsible for their generation makes these
particles and experiments searching for them a powerful window towards fundamental physics.

\begin{theacknowledgments}
The author thanks the organizers of the AXIONS 2010 for a wonderful meeting and Mark Goodsell and Andreas Ringwald for discussions.
\end{theacknowledgments}



\bibliographystyle{aipproc}   

\bibliography{sample}


\end{document}